\documentclass[12pt]{article}
\usepackage{amssymb,amsmath,epsfig}

\begin{document}
\title{\bf Effects of Electromagnetic Field on the Dynamical Instability of
Expansionfree Gravitational Collapse}

\author{M. Sharif$^1$ \thanks{msharif.math@pu.edu.pk} and M. Azam$^{1,2}$
\thanks{azammath@gmail.com}\\
$^1$ Department of Mathematics, University of the Punjab,\\
Quaid-e-Azam Campus, Lahore-54590, Pakistan.\\
$^2$ Division of Science and Technology, University of Education,\\
Township Campus, Lahore-54590, Pakistan.}

\date{}

\maketitle
\begin{abstract}
In this paper, we discuss the effects of electromagnetic field on
the dynamical instability of a spherically symmetric expansionfree
gravitational collapse. Darmois junction conditions are formulated
by matching interior spherically symmetric spacetime to exterior
Reissner-Nordstr$\ddot{o}$m spacetime. We investigate the role of
different terms in the dynamical equation at Newtonian and post
Newtonian regimes by using perturbation scheme. It is concluded that
instability range depends upon pressure anisotropy, radial profile
of energy density and electromagnetic field, but not on the
adiabatic index $\Gamma$. In particular, the electromagnetic field
reduces the unstable region.
\end{abstract}

\section{Introduction}

The stability/instability of self-gravitating objects has great
importance in general relativity. It is well-known that different
ranges of stability would imply different kinds of evolution in
the collapse as well as structure formation of self-gravitating
objects. The adiabatic index $\Gamma$ defines the range of
instability which is less than $\frac{4}{3}$ for a spherically
symmetric distribution of isotropic perfect fluid \cite{1}. Also,
it is obvious that a stellar model can exist only if it is stable
against fluctuations. A stable stationary black hole solution
under perturbations tells the final state of dynamical evolution
of a gravitating system.

The expansion scalar, $\Theta$, measures the rate at which small
volumes of the fluid may change. In the expanding sphere, the
increase in volume due to increasing area of external surface must
be reimbursed with the increasing area of internal boundary surface.
A similar behavior of surface area can be observed in the case of
contraction. Thus we have to keep $\Theta$ vanishing in each case.
Skripkin \cite{2} explored the central explosion of a spherically
symmetric fluid distribution with expansionfree scalar. This leads
to the formation of Minkowskian cavity at the center of the fluid.
Eardley and Smarr \cite{3} investigated that the collapse of
self-gravitating fluids would lead to formation of naked singularity
for inhomogeneous energy density but to black hole for homogenous
case. It was found that expansionfree model requires locally
anisotropic fluid and inhomogeneous energy density
\cite{4}-\cite{6}. Herrera et al. \cite{7} found that inhomogeneous
expansionfree dust models with negative energy density has no
physical significance.  The same authors \cite{8} discussed cavity
evolution in relativistic self-gravitating fluid.

Rosseland \cite{9} was the first to study self-gravitating
spherically symmetric charged fluid distribution. Since then many
people have considered the effect of electromagnetic charge on the
structure and evolution of self-gravitating systems
\cite{10}-\cite{14}. Di Prisco et al. \cite{15} explored the effect
of charge on the relation between the Weyl tensor and the
inhomogeneity of energy density and concluded that Coulomb repulsion
might prevent the gravitational collapse of the sphere. Thirukkanesh
and Maharaj \cite{16} investigated that gravitational attraction is
compensated by the Coulomb's repulsive force along with gradient
pressure in a gravitational collapse. Sharif and Abbas \cite{17}
discussed the effect of electromagnetic field on spherically
symmetric gravitational collapse with cosmological constant. Sharif
and Sundas \cite{18} used Misner-Sharp formalism to discuss charged
cylindrical collapse of anisotropic fluid and found that electric
charge increases the active gravitational mass.

It is evident that anisotropy, free streaming radiation, thermal
conduction and shearing viscosity affect the evolution of
self-gravitating systems. In literature \cite{19,20}, it is shown
that the thermal effects reduce the range of instability. Chan et
al. \cite{21} explored that the instability range depends upon the
local anisotropy of the unperturbed fluid. The same authors
\cite{22} found the effects of shearing viscous fluid on the
instability range. Chan \cite{23} studied collapsing radiating
star with shearing viscosity and concluded that it would increase
anisotropy of pressure as well as the value of effective adiabatic
index. Horvat et al. \cite{24} explored that instability of
anisotropic star occurs at higher surface compactness when the
anisotropy of the pressure is present. Herrera et al. \cite{25}
discussed the dynamical instability of expansionfree fluid at
Newtonian and post Newtonian order and found that the range of
instability is determined by the anisotropic pressure and radial
profile of the energy density. In a recent paper \cite{26}, this
problem has been explored in $f(R)$ gravity.

In this paper, we take spherically symmetric distribution of
collapsing fluid along with electromagnetic field and investigate
how electromagnetic field would affect the range of instability.
Darmois Junction conditions \cite{27} are used to match the
interior spherically symmetric spacetime to exterior
Reissner-Nordstr$\ddot{o}$m (RN) spacetime on the external
hypersurface and on the internal hypersurface Minkowski spacetime
within the cavity to the fluid distribution. We find that
electromagnetic field, energy density and anisotropic pressure
affect the stability of the system.

The paper has the following format. In section \textbf{2}, we
discuss Einstein-Maxwell equations and some basic properties of
anisotropic fluid. Section \textbf{3} provides the formulation of
junction conditions. In section \textbf{4}, the perturbation scheme
is applied on the field as well as dynamical equations. We discuss
the Newtonian and post Newtonian regimes and obtain the dynamical
equation in section \textbf{5}. Results are summarized in the last
section.

\section{Fluid Distribution and the Field Equations}

Consider a spherically symmetric distribution of charged collapsing
fluid bounded by a spherical surface $\Sigma$. The line element for
the interior region is the most general spherically symmetric metric
given by
\begin{equation}\label{1}
ds^2_-=-A^2(t,r)dt^{2}+B^2(t,r)dr^{2}+R^2(t,r)(d\theta^{2}
+\sin^2\theta{d\phi^2}),
\end{equation}
where we assume comoving coordinates inside the hypersurface
$\Sigma$. The interior coordinates are taken as
$x^{-0}=t,~x^{-1}=r,~x^{-2}=\theta,~x^{-3}=\phi$. It is assumed
that the fluid is locally anisotropic and the energy-momentum
tensor for such a fluid is given by
\begin{equation}\label{2}
T^-_{\alpha\beta}=(\mu+p_{\perp})u_{\alpha}u_{\beta}+p_{\perp}g_{\alpha\beta}+
(p_r-p_{\perp})\chi_{\alpha} \chi_{\beta},
\end{equation}
where $\mu$ is the energy density, $p_{\perp}$ the tangential
pressure, $p_r$ the radial pressure, $u_{\alpha}$ the four-velocity
of the fluid and $\chi_{\alpha}$ is the unit four-vector along the
radial direction. Using the following definitions in comoving
coordinates
\begin{equation}\label{3}
u^{\alpha}=A^{-1}\delta^{\alpha}_{0},\quad
\chi^{\alpha}=B^{-1}\delta^{\alpha}_{1},
\end{equation}
we can write
\begin{equation*}
u^{\alpha}u_{\alpha}=-1,\quad\chi^{\alpha}\chi_{\alpha}=1,\quad
\chi^{\alpha}u_{\alpha}=0.
\end{equation*}
The expansion scalar is defined as
\begin{equation}\label{4}
\Theta=u^{\alpha}_{;\alpha}=\frac{1}{A}\left(\frac{\dot{B}}{B}
+2\frac{\dot{R}}{R}\right).
\end{equation}
Here dot and prime represent derivatives with respect to $t$ and
$r$ respectively.

The Maxwell equations can be written as
\begin{eqnarray}\label{5}
F_{\alpha\beta}=\phi_{\beta,\alpha}-\phi_{\alpha,\beta}, \quad
F^{\alpha\beta}_{;\beta}=4{\pi}J^{\alpha},
\end{eqnarray}
where $\phi_\alpha$ is the four potential and $J^{\alpha}$ is the
four current. The electromagnetic energy-momentum tensor is given by
\begin{eqnarray}\label{7}
E_{\alpha\beta}=\frac{1}{4\pi}\left(F^\gamma_{\alpha}F_{\beta\gamma}
-\frac{1}{4}F^{\gamma\delta}F_{\gamma\delta}g_{\alpha\beta}\right),
\end{eqnarray}
where $F_{\alpha\beta}$ is the Maxwell field tensor. Since the
charge is at rest with respect to comoving coordinates, the magnetic
field will be zero. Thus we can write
\begin{eqnarray}\label{8}
\phi_{\alpha}=\left({\phi}(t,r),0,0,0\right),\quad
J^{\alpha}={\xi}u^{\alpha},
\end{eqnarray}
where $\xi$ is the charge density. The conservation of charge
requires
\begin{equation}\label{9}
q(r)=4\pi\int^r_{0}{\xi}B{R^2}dr
\end{equation}
which is the electric charge interior to radius $R$. Using
Eq.(\ref{1}), the Maxwell equations (\ref{5}) yield
\begin{eqnarray}\label{10}
{\phi''}-\left(\frac{A'}{A}+\frac{B'}{B}-2\frac{R'}{R}\right){\phi'}
&=&4\pi\xi{AB^{2}},\\\label{11}
\dot{\phi'}-\left(\frac{\dot{A}}{A}+\frac{\dot{B}}{B}
-2\frac{\dot{R}}{R}\right){\phi'}&=&0.
\end{eqnarray}
Integration of Eq.(\ref{10}) implies
\begin{eqnarray}\label{12}
{\phi'}=\frac{qAB}{R^{2}}.
\end{eqnarray}
The electric field intensity is defined as
\begin{eqnarray}\label{13}
E(t,r)=\frac{q}{4{\pi}R^{2}}.
\end{eqnarray}

The Einstein field equations
\begin{equation}\label{14}
G^-_{\alpha\beta}=8\pi\left(T^{-}_{\alpha\beta}
+E^-_{\alpha\beta}\right),
\end{equation}
for the interior metric gives the following set of equations
\begin{eqnarray}\nonumber
&&8{\pi}A^{2}({\mu}+2{\pi}E^2)=\left(\frac{2\dot{B}}{B}
+\frac{\dot{R}}{R}\right)\frac{\dot{R}}{R}\\\label{15}&-&\left(\frac{A}{B}\right)^2
\left[\frac{2R''}{R}+\left(\frac{R'}{R}\right)^2-\frac{2B'R'}{BR}
-\left(\frac{B}{R}\right)^2\right],\\\label{16}
&&0=-2\left(\frac{\dot{R'}}{R}-\frac{\dot{R}A'}{RA}
-\frac{\dot{B}R'}{BR}\right), \\\nonumber
&&8{\pi}B^{2}(p_{r}-2{\pi}E^2)=-\left(\frac{B}{A}\right)^2\left
[\frac{2\ddot{R}}{R}-\left(\frac{2\dot{A}}{A}
-\frac{\dot{R}}{R}\right) \frac{\dot{R}}{R}\right]\\\label{17}
&+&\left(\frac{2A'}{A}+\frac{R'}{R}\right)\frac{R'}{R}
-\left(\frac{B}{R}\right)^2, \\\nonumber
&&8{\pi}R^{2}(p_{\perp}+2{\pi}E^2)=8{\pi}R^{2}(p_{\perp}
+2{\pi}E^2)\sin^{-2}\theta\\\nonumber
&=&-\left(\frac{R}{A}\right)^2\left[\frac{\ddot{B}}{B}
+\frac{\ddot{R}}{R}-\frac{\dot{A}}{A}
\left(\frac{\dot{B}}{B}+\frac{\dot{R}}{R}\right)
+\frac{\dot{B}\dot{R}}{BR}\right]\\\label{18}
&+&\left(\frac{R}{B}\right)^2\left[\frac{A''}{A}
+\frac{R''}{R}-\frac{A'B'}{AB}\right.
\left.+\left(\frac{A'}{A}-\frac{B'}{B}\right)\frac{R'}{R}\right].
\end{eqnarray}
The mass function is defined as follows \cite{28}
\begin{equation}\label{19}
m(t,r)=\frac{R}{2}(1-g^{\alpha\beta}R_{,\alpha}R_{,\beta})
=\frac{R}{2}\left(1+\frac{\dot{R}^2}{A^2}
-\frac{R'^2}{B^2}\right)+\frac{q^2}{2R}.
\end{equation}
Differentiating this equation with respect to $r$ and using
Eq.(\ref{15}), we get
\begin{equation}\label{20}
m'=4\pi{\mu}R'R^2+16\pi^2R^2E(RE'+2R'E).
\end{equation}

The proper time and radial derivatives are given by
\begin{equation}\label{21}
D_{T}=\frac{1}{A}\frac{\partial}{\partial t},\quad
D_{R}=\frac{1}{R'}\frac{\partial}{\partial r},
\end{equation}
where $R$ is the areal radius of the spherical surface. The
velocity of the collapsing fluid is defined by the proper time
derivative of $R$, i.e.,
\begin{equation}\label{22}
U=D_{T}R=\frac{\dot{R}}{A}
\end{equation}
which is always negative in case of collapse. Using this
expression, Eq.(\ref{19}) can be written as
\begin{equation}\label{23}
\tilde{E}\equiv\frac{R'}{B}=\left[1+U^{2}-\frac{2m}{R}
+\left(\frac{q}{R}\right)^2\right]^{1/2}.
\end{equation}
The conservation of energy-momentum tensor yields
\begin{equation}\label{24}
(T^{-\alpha\beta}+E^{-\alpha\beta})_{;\beta}u_\alpha=-\frac{1}{A}
\left[\dot{\mu}+(\mu+p_r)\frac{\dot{B}}{B}
+2(\mu+p_\perp)\frac{\dot{R}}{R}\right]=0
\end{equation}
which can be rewritten as
\begin{equation}\label{25}
\dot{\mu}+(\mu+p_r)A\Theta+2(p_\perp-p_r)\frac{\dot{R}}{R}=0,
\end{equation}
and
\begin{eqnarray}\nonumber
(T^{-\alpha\beta}+E^{-\alpha\beta})_{;\beta}\chi_{\alpha}&=&\frac{1}{B}
\left[p'_{r}+(\mu+p_r)\frac{A'}{A}+2(p_{r}-p_\perp)\frac{R'}{R}
\right.\\\label{26}&-&\left.\frac{E}{R}(4{\pi}RE'+8{\pi}R'E)\right]=0.
\end{eqnarray}

\section{Junction Conditions}

In this section, we formulate the Darmois junction conditions for
the general spherically symmetric spacetime in the interior region
and RN spacetime in the exterior region. The line element for RN
spacetime in Eddington-Finkelstein coordinates is given as
\begin{equation}\label{27}
ds^2_+=-\left(1-\frac{2M}{\rho}+\frac{Q^2}{\rho^2}\right)d\nu^2
-2d{\rho}d{\nu}+\rho^2(d\theta^2+\sin^2{\theta}d\phi^2),
\end{equation}
where $M$, $Q$ and $\nu$ are the total mass, charge and retarded
time respectively. For smooth matching of the interior and exterior
regions, Darmois conditions \cite{27} can be stated as follows:\\\\
1. The continuity of the line elements over $\Sigma$
\begin{equation}\label{28}
\left(ds^{2}_{-}\right)_{\Sigma}=\left(ds^{2}_{+}\right)_{\Sigma}
=\left(ds^{2}\right)_{\Sigma}.
\end{equation}
2. The continuity of the extrinsic curvature over $\Sigma$
\begin{equation}\label{29}
\left[K_{ij}\right]=K^{+}_{ij}-K^{-}_{ij}=0,\quad (i,j=0,2,3).
\end{equation}
The boundary surface $\Sigma$ in terms of interior and exterior
coordinates can be defined as
\begin{eqnarray}\label{30}
f_{-}(t,r)&=&r-r_{\Sigma}=0,\\\label{31}
f_{+}(\nu,\rho)&=&\rho-\rho(\nu_{\Sigma})=0,
\end{eqnarray}
where $r_{\Sigma}$ is a constant. Using Eqs.(\ref{30}) and
(\ref{31}), the interior and exterior metrics take the following
form over $\Sigma$
\begin{eqnarray}\label{32}
(ds^2_{-})_{\Sigma}&=&-A^2(t,r_{\Sigma})dt^{2}+R^2(t,r_{\Sigma})(d\theta^{2}
+\sin^2\theta{d\phi^{2}}),\\\label{33}
(ds^2_{+})_{\Sigma}&=&-\left(1-\frac{2M}{\rho_{\Sigma}}
+\frac{Q^2}{\rho^2_{\Sigma}}+2\frac{d\rho_{\Sigma}}{d\nu}\right)d\nu^2
+\rho^2_{\Sigma}(d\theta^2+\sin^2\theta{d\phi^2}).
\end{eqnarray}

The continuity of the first fundamental form implies
\begin{eqnarray}\label{34}
\frac{dt}{d\tau}&=&A(t, r_{\Sigma})^{-1},\quad
R(t,r_{\Sigma})=\rho_{\Sigma}(\nu),\\\label{35}
\left(\frac{d\nu}{d\tau}\right)^{-2}&=&\left(1-\frac{2M}{\rho_{\Sigma}}
+\frac{Q^2}{\rho^2_{\Sigma}}+2\frac{d\rho_{\Sigma}}{d\nu}\right).
\end{eqnarray}
For the second fundamental form, we evaluate outward unit normals
to $\Sigma$ by using Eqs.(\ref{30}) and (\ref{31}) as follows
\begin{eqnarray}\label{36}
n^{-}_{\alpha}&=&\left(0,B(t,r_{\Sigma}),0,0\right),\\\label{37}
n^{+}_{\alpha}&=&\left(1-\frac{2M}{\rho_{\Sigma}}
+\frac{Q^2}{\rho^2_{\Sigma}}+2\frac{d\rho_{\Sigma}}{d\nu}\right)^{-\frac{1}{2}}
\left(-\frac{d\rho_{\Sigma}}{d\nu},1,0,0\right).
\end{eqnarray}
The non-vanishing components of the extrinsic curvature in terms
of interior and exterior coordinates are
\begin{eqnarray}\label{38}
&&K^{-}_{00}=-\left[\frac{A'}{AB}\right]_{\Sigma},\quad
K^{-}_{22}=\left[\frac{RR'}{B}\right]_{\Sigma},\quad
K^{-}_{33}=K^{-}_{22}\sin^{2}\theta, \\\label{39}
&&K^{+}_{00}=\left[\left(\frac{d^{2}\nu}{d\tau^{2}}\right)
\left(\frac{d\nu}{d\tau}\right)^{-1}
-\left(\frac{d\nu}{d\tau}\right)\left(\frac{M}{\rho^{2}}
-\frac{Q^{2}}{\rho^{3}}\right)\right]_{\Sigma},\\\label{40}
&&K^{+}_{22}=\left[\left(\frac{d\nu}{d\tau}\right)
\left(1-\frac{2M}{r}-\frac{Q^2}{r^2}\right)r
+\left(\frac{dr}{d\tau}\right)r\right]_{\Sigma},\\\label{41}
&&K^{+}_{33}=K^{+}_{22}\sin^{2}\theta.
\end{eqnarray}
Making use of Eqs.(\ref{29}), (\ref{34}) and (\ref{35}), we get
\begin{eqnarray}\label{42}
M\overset{\Sigma}=m(t,r)\quad \Longleftrightarrow\quad
q(r)\overset{\Sigma}=Q
\end{eqnarray}
and
\begin{eqnarray}\nonumber
&&2\left(\frac{\dot{R'}}{R}-\frac{\dot{R}A'}{RA}-\frac{\dot{B}R'}{BR}\right)
\overset{\Sigma}=-\frac{B}{A}\left[\frac{2\ddot{R}}{R}-\left(\frac{2\dot{A}}{A}-
\frac{\dot{R}}{R}\right) \frac{\dot{R}}{R}\right]\\\label{43}
&&+\frac{A}{B}\left[\left(\frac{2A'}{A}+\frac{R'}{R}\right)\frac{R'}{R}
-\left(\frac{B}{R}\right)^2\right],
\end{eqnarray}
where $q(r)=Q$ has been used. Comparing Eq.(\ref{43}) with
Eqs.(\ref{16}) and (\ref{17}), we obtain
\begin{equation}\label{44}
p_{r}\overset{\Sigma}=0.
\end{equation}

The expansionfree models require the existence of internal vacuum
cavity within the fluid distribution. The matching of Minkowski
spacetime within cavity to the fluid distribution on $\Sigma^{(i)}$
(boundary surface between cavity and fluid) gives
\begin{equation}\label{45}
m(t,r)\overset{\Sigma^{(i)}}{=}0,\quad
p_{r}\overset{\Sigma^{(i)}}{=}0.
\end{equation}

\section{The Perturbation Scheme}

This section is devoted to perturb the field equations, Bianchi
identities and all the material quantities by using the perturbation
scheme \cite{19,20} upto first order. Initially, all the quantities
have only radial dependence, i.e., fluid is in static equilibrium.
After that, all the quantities and the metric functions have time
dependence as well in their perturbation. These are given by
\begin{eqnarray}\label{46}
A(t,r)&=&A_0(r)+\lambda T(t)a(r),\\\label{47}
B(t,r)&=&B_0(r)+\lambda T(t)b(r),\\\label{48}
R(t,r)&=&R_0(r)+\lambda T(t)c(r),\\\label{49}
E(t,r)&=&E_0(r)+\lambda T(t)e(r),\\\label{50}
\mu(t,r)&=&\mu_0(r)+\lambda {\bar{\mu}}(t,r),\\\label{51}
p_r(t,r)&=&p_{r0}(r)+\lambda {\bar{p_r}}(t,r),\\\label{52}
p_{\perp}(t,r)&=&p_{\perp0}(r)+\lambda{\bar{p_{\perp}}}(t,r),\\\label{53}
m(t,r)&=&m_0(r)+\lambda{\bar{m}}(t,r),\\\label{54}
\Theta(t,r)&=&\lambda {\bar{\Theta}}(t,r),
\end{eqnarray}
where $0<\lambda\ll1$. By the freedom allowed in radial
coordinates, we choose $R_0(r)=r$. The static configuration
(unperturbed) of Eqs.(\ref{15})-(\ref{18}) is obtained by using
Eqs.(\ref{46})-(\ref{52}) as follows
\begin{eqnarray}\label{55}
8{\pi}\left(\mu_{0}+2{\pi}E^{2}_{0}\right)
=\frac{1}{(B_0r)^2}\left(2r\frac{B_0'}{B_0}+B_0^2-1\right),\\\label{56}
8{\pi}\left(p_{r0}-2{\pi}E^{2}_{0}\right)
=\frac{1}{(B_0r)^2}\left(2r\frac{A_0'}{A_0}-B_0^2+1\right),\\\label{57}
8{\pi}\left(p_{\perp0}+2{\pi}E^{2}_{0}\right)
=\frac{1}{B_0^2}\left[\frac{A_0''}{A_0}-\frac{A_0'}{A_0}\frac{B_0'}{B_0}
+\frac{1}{r}\left(\frac{A_0'}{A_0}-\frac{B_0'}{B_0}\right)\right].
\end{eqnarray}

The corresponding perturbed field equations become
\begin{eqnarray}\nonumber
8{\pi}{\bar\mu}+32{\pi}^2{E_0}Te&=&-\frac{2T}{B_0^2}
\left[\left(\frac{c}{r}\right)''-\frac{1}{r}
\left(\frac{b}{B_0}\right)'
-\left(\frac{B_0'}{B_0}-\frac{3}{r}\right)
\left(\frac{c}{r}\right)'\right.\\\label{58}
&-&\left.\left(\frac{b}{B_0}-\frac{c}{r}\right)
\left(\frac{B_0}{r}\right)^2\right]-16{\pi}\frac{Tb}{B_{0}}
\left({\mu_{0}}+2{\pi}E^{2}_{0}\right),\\\label{59}
0&=&2\frac{\dot{T}}{A_0B_0}\left[\left(\frac{c}{r}\right)'
-\frac{b}{rB_0}-\left(\frac{A'_0}{A_0}-\frac{1}{r}\right)\frac{c}{r}\right],
\end{eqnarray}
\begin{eqnarray}\nonumber
8{\pi}{\bar{p_{r}}}-32{\pi}^2{E_0}Te&=&-\frac{2\ddot{T}}{A_0^2}\frac{c}{r}+
\frac{2T}{rB_0^2}\left[\left(\frac{a}{A_0}\right)'+
\left(r\frac{A_0'}{A_0}+1\right)\left(\frac{c}{r}\right)'\right.\\\label{60}
&-&\left.\frac{B_0^2}{r}\left(\frac{b}{B_0}-\frac{c}{r}\right)\right]
-16{\pi}\frac{Tb}{B_0}\left({p_{r0}}-2{\pi}E^{2}_{0}\right),\\\nonumber
8{\pi}{\bar{p_{\perp}}}+32{\pi}^2{E_0}Te&=&-\frac{\ddot{T}}{A_0^2}
\left[\frac{b}{B_0}+\frac{c}{r}\right]+\frac{T}{B_0^2}
\left[\left(\frac{a}{A_0}\right)''
+\left(\frac{c}{r}\right)''\right.\\\nonumber
&+&\left.\left(\frac{2A_0'}{A_0}-\frac{B_0'}{B_0}+\frac{1}{r}\right)
\left(\frac{a}{A_0}\right)'-
\left(\frac{A_0'}{A_0}+\frac{1}{r}\right)
\right.\\\nonumber&\times&\left.\left(\frac{b}{B_0}\right)'
+\left(\frac{A_0'}{A_0} -\frac{B_0'}{B_0}+\frac{2}{r}\right)
\left(\frac{c}{r}\right)'\right]\\\label{61}
&-&16{\pi}\frac{Tb}{B_0}\left({p_{\perp0}}+2{\pi}E^{2}_{0}\right).
\end{eqnarray}
The Bianchi identities (\ref{24}) and (\ref{26}) for the static
configuration yields
\begin{eqnarray}\label{62}
\frac{1}{B_0}\left[p_{r0}'+(\mu_0+p_{r0})\frac{A_0'}{A_0}
+\frac{2}{r}(p_{r0}-p_{\perp0})\right]
-\frac{4\pi{E_0}}{B_0{r}}\left[2E_0+rE'_{0}\right]=0,
\end{eqnarray}
which can be rewritten as
\begin{eqnarray}\label{63}
\frac{A'_0}{A_0}&=&-\frac{1}{\mu_0+p_{r0}}
\left[p'_{r0}+\frac{2}{r}(p_{r0}-p_{\perp0})-
\frac{4{\pi}E_0}{r}(2E_0+rE'_0)\right].
\end{eqnarray}
The perturbed configurations imply
\begin{eqnarray}\label{64}
&&\frac{1}{A_0}\left[\dot{\bar{\mu}}+(\mu_0+p_{r0})\dot{T}\frac{b}{B_0}
+2(\mu_0+p_{\perp0})\dot{T}\frac{c}{r}\right]=0, \\\nonumber
&&\frac{1}{B_0}\left[\bar{p'_r}+(\mu_0+p_{r0}){T}\left(\frac{a}{A_0}\right)'
+(\bar{\mu}+\bar{p_r})\frac{A'_0}{A_0} \right.\\\nonumber
&+&\left.2(p_{r0}-p_{\perp0}){T}\left(\frac{c}{r}\right)'+2(\bar{p_r}
-\bar{p_\perp})\frac{1}{r}\right]\\\label{65}
&-&\frac{4\pi{E_0}T}{{B_0}r}\left(4e+2r{E_0}\left(\frac{c}{r}\right)'
+re'+re\frac{E'_0}{E_0}\right)=0.
\end{eqnarray}
Integration of Eq.(\ref{64}) yields
\begin{eqnarray}\label{66}
\bar\mu=-\left[(\mu_0+p_{r0})\frac{b}{B_0}+2(\mu_0+p_{\perp0})\frac{c}{r}\right]T.
\end{eqnarray}

The expansion scalar turns out to be
\begin{eqnarray}\label{67}
\bar\Theta&=&\frac{\dot{T}}{A_0}\left(\frac{b}{B_0}+\frac{2c}{r}\right).
\end{eqnarray}
Using expansionfree condition, it follows
\begin{equation}\label{68}
\frac{b}{B_0}=-2\frac{c}{r}.
\end{equation}
Inserting this value in Eq.(\ref{59}), we obtain
\begin{equation}\label{69}
c=k\frac{A_0}{r^2},
\end{equation}
where $k$ is an integration constant. Using Eq.(\ref{68}) in
(\ref{66}), we get
\begin{equation}\label{70}
\bar{\mu}=2(p_{r0}-p_{\perp0})T\frac{c}{r}.
\end{equation}
This shows that perturbed energy density comes from the static
configuration of pressure anisotropy. Similarly, the unperturbed and
perturbed configuration for Eq.(\ref{19}) lead to
\begin{eqnarray}\label{71}
m_0&=&\frac{r}{2}\left(1-\frac{1}{B_0^2}\right)
+8\pi^2{E^2_0}r^3,\\\label{72}
\bar{m}&=&-\frac{T}{B_0^2}\left[r\left(c'-\frac{b}{B_0}\right)
+(1-B_0^2)\frac{c}{2}\right]
+8\pi^2E_0{T}\left(2r^3+3r^2c{E_0}\right).
\end{eqnarray}
Using the matching condition (\ref{44}), Eq.(\ref{51}) implies
\begin{equation}\label{73}
p_{r0}\overset{\Sigma}=0,\quad \bar{p}_{r}\overset{\Sigma}=0.
\end{equation}
Inserting these values in Eq.(\ref{60}), we obtain
\begin{equation}\label{74}
\ddot{T}(t)-\alpha(r){T}(t)\overset{\Sigma}=0,
\end{equation}
where
\begin{eqnarray}\nonumber
\alpha(r)&\overset{\Sigma}=&\left(\frac{A_{0}}{B_{0}}\right)^2
\left[\left(\frac{a}{A_0}\right)'+\left(r\frac{A_0'}{A_0}+1\right)
\left(\frac{c}{r}\right)'\right.\\\label{75}
&-&\left.\frac{B_0^2}{r}\left(\frac{b}{B_0}-\frac{c}{r}\right)
+16{\pi}^2rE_{0}B_{0}\left(eB_0+bE_{0}\right)\right]\frac{1}{c}.
\end{eqnarray}
In order to explore instability region, all the functions involved
in the above equation are taken such that $\alpha_{\Sigma}$ is
positive. The corresponding solution of Eq.(\ref{74}) is given by
\begin{equation}\label{76}
T(t)=-\exp(\sqrt{\alpha_{\Sigma}}t).
\end{equation}
This shows that the system starts collapsing at $t=-\infty$ with
$T(-\infty)=0$ keeping it in static position. It goes on collapsing
with the increase of $t$.

\section{Newtonian and Post Newtonian Terms and Dynamical Instability}

This section investigates the terms corresponding to Newtonian
(N), post Newtonian (pN) and post post Newtonian (ppN) regimes.
This is done by converting relativistic units into c.g.s. units
and expanding upto order $c^{-4}$ in the dynamical equation. For
the N approximation, it is assumed that
$$\mu_0\gg p_{r0},\quad\mu_0\gg p_{\perp0}.$$
For the metric coefficients expanded upto pN approximation, we
take
\begin{equation}\label{77}
A_0=1-\frac{Gm_0}{c^2r},\quad B_0=1+\frac{Gm_0}{c^2r},
\end{equation}
where $G$ is the gravitational constant and $c$ is the speed of
light. Using Eqs.(\ref{56}) and (\ref{71}), it follows that
\begin{equation}\label{78}
\frac{A_0'}{A_0}=\frac{8\pi{p_{r0}}r^3+2m_0-32{\pi^2}{E^2_0}r^3}
{2r(r-2{m_0}+16{\pi^2}{E^2_0}r^3)},
\end{equation}
which together with Eq.(\ref{62}) leads to
\begin{eqnarray}\label{79}
p_{r0}'&=&-\left[\frac{8\pi{p_{r0}}r^3+2m_0-32{\pi^2}
{E^2_0}r^3}{2r(r-2{m_0}+16{\pi^2}{E^2_0}r^3)}\right](\mu_0+p_{r0})
\\\nonumber&+&\frac{2}{r}
(p_{\perp0}-p_{r0})+\frac{4{\pi}E_0}{r}(2E_0+rE'_0).
\end{eqnarray}

In view of dimensional analysis, this equation can be written in
c.g.s. units as follows
\begin{eqnarray}\label{80}
p_{r0}'&=&-G\left[\frac{c^{-2}8\pi{p_{r0}}r^3+2m_0-32c^{-2}{\pi^2}
{E^2_0}r^3}{2r(r-2Gc^{-2}{m_0}+16Gc^{-4}{\pi^2}{E^2_0}r^3)}\right]
(\mu_0+c^{-2}p_{r0})\\\nonumber &+&\frac{2}{r}
(p_{\perp0}-p_{r0})+\frac{4{\pi}E_0}{r}(2E_0+rE'_0).
\end{eqnarray}
When we expand this equation upto $c^{-4}$ order and rearrange
lengthy calculations, we have
\begin{eqnarray}\label{81}
p_{r0}'&=&-G\frac{\mu_0m_0}{r^2}+\frac{2}{r}(p_{\perp0}-p_{r0})+
\frac{4\pi}{r}\left(2E^2_0+r{E_0}E'_0\right)\\\nonumber&-&
\frac{G}{c^{2}r^3}\left(2G{\mu_0}m^{2}_0+p_{r0}m_{0}r+4{\pi}
\mu_{0}p_{r0}r^4-16\pi^2{E^2_0}{\mu_0}r^4\right)\\\nonumber
&-&\frac{G}{c^{4}r^4}\left(4G^2{\mu_0}m^{2}_0
+2Gp_{r0}m^{2}_{0}r+4{\pi}\mu_{0}p_{r0}r^4\right.\\\nonumber
&-&\left.32\pi^2G{E^2_0}m_0{\mu_0}r^4
-16\pi^2{E^2_0}{p_{r0}}r^5\right).
\end{eqnarray}
Here the terms with coefficient $c^0$ are called N order terms,
coefficient with $c^{-2}$ of pN order and with $c^{-4}$ are of ppN
order terms. The relationship between $\bar{\mu}$ and $\bar{p}_r$
is given by \cite{19,20}
\begin{equation}\label{82}
\bar{p}_r=\Gamma\frac{p_{r0}}{\mu_0+p_{r0}}\bar{\mu}.
\end{equation}
It is noted that the fluid under the expansionfree condition
evolves without being compressed \cite{29}. Thus the adiabatic
index $\Gamma$ (which measures the variation of pressure for a
given variation of density) is irrelevant here for the case of
expansionfree evolution as the perturbed energy density depends on
the static configuration. Using Eq.(\ref{70}) in the above
equation, it follows that
\begin{equation}\label{83}
\bar{p_r}=2\Gamma\frac{p_{r0}}{\mu_0+p_{r0}}(p_{r0}-p_{\perp0})T\frac{c}{r}.
\end{equation}
From Eqs.(\ref{55}) and (\ref{71}), we get
\begin{eqnarray}\label{84}
\frac{B_0'}{B_0}&=&\frac{8\pi{\mu_{0}}r^3-2m_0-32{\pi^2}{E^2_0}r^3}
{2r(r-2{m_0}+16{\pi^2}{E^2_0}r^3)}.
\end{eqnarray}

Next, we develop dynamical equation by substituting Eq.(\ref{61})
along with Eqs.(\ref{68}) and (\ref{76}) in Eq.(\ref{65}) and
neglecting the ppN order terms $\bar{p_r},~\bar{\mu}
\frac{A'_0}{A_0}$, it follows that
\begin{eqnarray}\nonumber
&&8\pi(\mu_0+p_{r0})r\left(\frac{a}{A_0}\right)'
+16\pi(p_{r0}-p_{\perp0})r\left(\frac{c}{r}\right)'\\\nonumber
&-&64\pi({p_{\perp0}+2\pi{E^2_0}})\frac{c}{r}
-32\pi^{2}{E_0}\left(2e+2r{E_0}\left(\frac{c}{r}\right)'
+re'+re\frac{E'_0}{E_0}\right)\\\nonumber
&-&\frac{2}{B^2_{0}}\left[\left(\frac{a}{A_0}\right)''
+\left(\frac{c}{r}\right)''+\left(2\frac{A'_0}{A_0}-\frac{B_0'}{B_0}
+\frac{1}{r}\right)\left(\frac{a}{A_0}\right)'\right.\\\label{85}
&+&\left.\left(3\frac{A'_0}{A_0}-\frac{B_0'}{B_0}
+\frac{4}{r}\right)\left(\frac{c}{r}\right)'\right]
-2\frac{\alpha_{\Sigma}}{A^2_0}\frac{c}{r}=0.
\end{eqnarray}
In order to discuss instability conditions of this equation upto
pN order, we evaluate the following terms of dynamical equation.
Under expansionfree condition, Eq.(\ref{60}) can be written as
\begin{eqnarray}\nonumber
\left(\frac{a}{A_0}\right)'&=&-\frac{k{A_0}}{r^2}\left[16\pi(p_{r0}
-2\pi{E^2_0})B^2_0-\alpha_{\Sigma}\left(\frac{B_0}{A_0}\right)^2
\right.\\\label{86}&+&\left.\left(\frac{A'_0}{A_0}\right)^2
-\frac{2}{r}\frac{A'_0}{A_0}+\frac{3}{r^2}(B^2_0-1)\right]
-16{\pi}^2er{B^2_0}{E_0},
\end{eqnarray}
where Eqs.(\ref{69}) and (\ref{74}) has been used. We can write
two more equations by using Eqs.(\ref{86}) and (\ref{69}) as
follows
\begin{eqnarray}\nonumber
&&\left(\frac{a}{A_0}\right)''+\left(2\frac{A'_0}{A_0}-\frac{B'_0}{B_0}
+\frac{1}{r}\right)\left(\frac{a}{A_0}\right)'\\\nonumber
&=&k\frac{A_0}{r^2}
\left[16\pi{p_{r0}}{B^2_0}\left(\frac{1}{r}-\frac{B'_0}{B_0}\right)
-16\pi{p_{r0}}'{B^2_0}+\frac{2}{r}\left(\frac{A'_0}{A_0}\right)'
\right.\\\nonumber
&-&\left.\frac{2}{r}\frac{A'_0}{A_0}\frac{B'_0}{B_0}
+\frac{1}{r^2}\frac{A'_0}{A_0}(5-9B^2_0)-\frac{3}{r^2}\frac{B'_0}{B_0}(B^2_0+1)
+\frac{9}{r^3}(B^2_0-1)\right]\\\nonumber&+&\alpha_{\Sigma}k
\frac{A_0}{r^2}\left(\frac{B_0}{A_0}\right)^2\left(\frac{A'_0}{A_0}
-\frac{B'_0}{B_0}+\frac{1}{r}\right)-16\frac{kA_0{\pi}^2}{r^2}
\left[12E^2_{0}\frac{A'_0}{A_0}\right.\\\nonumber&+&\left.2{E^2_0}
\left(2\frac{B'_0}{B_0}-\frac{1}{r}\right)+8E_0E'_0\right]+32{\pi^2}r
\left(3eE_0\frac{B'_0}{B_0}+eE'_0+e'E_0\right)\\\label{87}
&-&64{\pi^2}erE_0\frac{A'_0}{A_0}. \\\nonumber
&&\left(\frac{c}{r}\right)''+\left(3\frac{A'_0}{A_0}-\frac{B'_0}{B_0}
+\frac{4}{r}\right)\left(\frac{c}{r}\right)'\\\label{88}&=&k\frac{A_0}{r^3}
\left[\left(\frac{A'_0}{A_0}\right)'-\frac{A'_0}{A_0}\frac{B'_0}{B_0}
-\frac{11}{r}\frac{A'_0}{A_0} +\frac{3}{r}\frac{B'_0}{B_0}\right].
\end{eqnarray}
Combining Eqs.(\ref{87}) and (\ref{88}), it follows that
\begin{eqnarray}\nonumber
&-&\frac{2}{B^2_0}\left[\left(\frac{a}{A_0}\right)''+\left(\frac{c}{r}\right)''
+\left(2\frac{A'_0}{A_0}-\frac{B'_0}{B_0}
+\frac{1}{r}\right)\left(\frac{a}{A_0}\right)'\right.\\\nonumber&-&\left.
\left(3\frac{A'_0}{A_0}-\frac{B'_0}{B_0}
+\frac{4}{r}\right)\left(\frac{c}{r}\right)'\right]-
2\frac{\alpha_{\Sigma}}{A^2_0}\frac{c}{r}
\end{eqnarray}
\begin{eqnarray}\nonumber
&=&32{\pi}k\frac{A_0}{r^2}
\left[p'_{r0}+p_{r0}\left(\frac{B'_{0}}{B_{0}}-\frac{1}{r}\right)\right]
-6k\frac{A_0}{B^2_{0}r^3}\left[\left(\frac{A'_0}{A_0}\right)'
\right.\\\nonumber&-&\left.\frac{A'_0}{A_0}\frac{B'_0}{B_0}-(3B^2_{0}+2)
\frac{1}{r}\frac{A'_0}{A_0}-\frac{1}{r}B_{0}B'_{0}+\frac{3}{r^2}(B^2_{0}-1)\right]
\\\nonumber&-&2\frac{\alpha_{\Sigma}k}{A_{0}r^2}
\left(\frac{A'_0}{A_0}+\frac{B'_0}{B_0}\right)-16\frac{kA_0{\pi}^2}{r^2}
\left[12E^2_{0}\frac{A'_0}{A_0}+2{E^2_0}\left(2\frac{B'_0}{B_0}-\frac{1}{r}\right)
\right.\\\label{89}&+&\left.8E_0E'_0\right]+32{\pi^2}r
\left(3eE_0\frac{B'_0}{B_0}+eE'_0+e'E_0\right)
-16{\pi^2}erE_0\frac{A'_0}{A_0}.
\end{eqnarray}

Inserting Eqs.(\ref{86}) and (\ref{89}) in Eq.(\ref{85}) and making
use of Eqs.(\ref{63}), (\ref{77}) and (\ref{84}), we obtain
dynamical equation at pN order (with $c = G = 1$)
\begin{eqnarray}\nonumber
&&-\frac{8\pi}{r^2}\left\{\left(1-\frac{m_0}{r}\right)\left[2p'_{ro}
+\frac{2}{r}(5p_{r0}-p_{\perp0})-8\pi{E_0E'_0}
-\frac{24\pi{E^2_0}}{r}\right]\right.\\\nonumber&+&\left.4\pi{\mu_0}r(4p_{r0}-{E^2_0})+
(\mu_0+p_{r0})r\left[\frac{3}{r^2}(B^2_0-1)-{\alpha_\Sigma}
\left(\frac{B_0}{A_0}\right)^2\right]\right\}\\\nonumber
&+&(\mu_0+p_{0})r\left[6\frac{m_0}{r^3}-\alpha_{\Sigma}
\left(1+3\frac{m_0}{r}\right)\right]-{\mu_0}r\left(\frac{3}{r^2}
+4\alpha_{\Sigma}\right)\left(\frac{m_0}{r}\right)^2\\\nonumber
&+&\frac{32{\pi}k}{r^2}\left[p'_{r0}-\frac{p_{r0}}{r}
+4{\pi}r{\mu_{0}}p_{r0}-\frac{m_0}{r}{p'_{r0}}-16{\pi}^2E^2_0r^2p_{r0}+
\frac{m_0}{r}p_{r0}\right.\\\nonumber&\times&\left.
(64{\pi}^3E^2r^3{\mu_0}+16{\pi}^2E^2_0r^2)\right]
-6\frac{k}{r^3}\left\{\frac{\mu'_{0}}{\mu^2_0}\left[p'_{r0}
+\frac{2}{r}(p_{r0}-p_{\perp0})\right.\right.\\\nonumber
&-&\left.\left.\frac{4{\pi}E_0}{r}(2E_0+rE'_0)\right]\right.
-\left(1-3\frac{m_0}{r}\right)
\frac{1}{\mu_0}\left[p''_{r0}-\frac{2}{r^2}(p_{r0}-p_{\perp0})
\right.\\\nonumber&+&\left.\frac{2}{r}(p'_{r0}-p'_{\perp0})-
\frac{4\pi}{r}(3E_0E'_0+rE_0E''_0+E^{'2}_0)
+\frac{4{\pi}E_0}{r^2}(2E_0+rE'_0)\right]\\\nonumber
&+&\frac{8\pi{r^3}\mu_0-2{m_0}-32{\pi}^2E^2_0r^3}{2r^2{\mu_0}}\left[p'_{r0}
+\frac{2}{r}(p_{r0}-p_{\perp0})-\frac{4{\pi}E_0}{r}(2E_0\right.
\\\nonumber&+&\left.rE'_0)\right]+\left(5-9\frac{m_0}{r}\right)
\frac{1}{r\mu_0}\left[p'_{r0}+\frac{2}{r^2}(p_{r0}-p_{\perp0})
-\frac{4{\pi}E_0}{r}(2E_0+rE'_0)\right]\\\nonumber
&-&{4\pi}{\mu_0}\left[1+\frac{m_0}{r}+2
\left(\frac{m_0}{r}\right)^2\right]+\frac{1}{r^2}\left[7\frac{m_0}{r}-5
\left(\frac{m_0}{r}\right)^2\right]+16{\pi^2}E^2_0\\\nonumber
&+&\left.64{\pi^3}\mu_0{E^2_0}r^2+48{\pi^2}E^2_0\frac{m_0}{r}\right\}
-2\frac{\alpha_{\Sigma}k}{r^2} \left\{-\left(1+\frac{m_0}{r}\right)
\frac{1}{r\mu_0}\left[p'_{r0}\right.\right.
\end{eqnarray}
\begin{eqnarray}\nonumber
&+&\left.\left.\frac{2}{r}(p_{r0}-p_{\perp0})-
\frac{4{\pi}E_0}{r}(2E_0+rE'_0)\right]+{4\pi}\left[1+3\frac{m_0}{r}+
\left(\frac{m_0}{r}\right)^2\right]r\mu_{0}\right.\\\nonumber
&-&\left.\frac{1}{r}\left[\frac{m_0}{r}+3
\left(\frac{m_0}{r}\right)^2\right]-\left(1+\frac{m_0}{r}\right)
64{\pi}^3E^2_0{\mu_0}r^3-32{\pi}^2m_0E^2_0\right\}\\\label{90}
&-&16\frac{kA_0{\pi}^2}{r^2}
\left[16\pi{\mu_0}E^2_0r^2-\frac{2E^2_0}{r}\left(1-\frac{m_0}{r}\right)
+8E_0E'_0\right]=0.
\end{eqnarray}
Using the fact that $\mu_0\gg p_{r0}$,  we discard the terms
belonging to pN and ppN order like $\frac{p_{r0}}{\mu_0}$ in the
above equation to obtain dynamical equation at N approximation as
follows
\begin{eqnarray}\nonumber
&&24\pi{\mu_0}+8\pi{\mid{p'_{r0}}\mid}r+2\left(\alpha_{\Sigma}
-\frac{21}{r^2}\right)\frac{m_0}{r}+\frac{32{\pi}^2r^3{E_0}}{k}
\left(2e+re'+re\frac{E'_0}{E_0}\right)\\\label{91}
&&=32\pi(5p_{r0}-2p_{\perp0})-416{\pi}^2{E^2_0} -32{\pi^2}rE_0E'_0.
\end{eqnarray}
Here $p'_{r0}<0$ shows that pressure is decreasing during collapse
of expansionfree fluid. Using Eq.(\ref{20}) in the above equation,
we get
\begin{eqnarray}\nonumber
&&\frac{4\pi}{9}{\mid{p'_{r0}}\mid}+\alpha_{\Sigma}\frac{{m_0}r^2}{9}
+\frac{16{\pi}^2r^6{E_0}}{9k}
\left(2e+re'+re\frac{E'_0}{E_0}\right)\\\nonumber
&=&\frac{16\pi}{9}(5p_{r0}-2p_{\perp0})r^3
+\frac{4\pi}{3}\left(7\int^{r}_{r_{{\Sigma}^{(i)}}}{\mu_0}r^2dr-{\mu_0}r^3\right)\\\label{92}
&+&\frac{16\pi^2}{3}\left[7\int^{r}_{r_{{\Sigma}^{(i)}}}({E_0E'_0}r^3+2E^2_0r^2)dr
-\frac{1}{3}(13{E^2_0}r^3-r^4E_0E'_0)\right].
\end{eqnarray}

For the instability of expansionfree fluid, we require that each
term in Eq.(\ref{92}) must be positive. For this purpose, the
positivity of the first term of Eq.(\ref{92}) leads to
$p_{r0}>(\frac{2}{5})p_{\perp0}$ and the positivity of the last two
terms is determined by considering the radial profile of the energy
density and electromagnetic field in the form $\mu_0={\gamma}r^m$
and $E_0={\delta}r^n$ respectively. Here $\gamma,~\delta$ are the
positive constants and $m,~n$ are constants defined in the interval
$(-\infty,\infty)$. Using these solutions, the last two terms of
Eq.(\ref{92}) will be positive for $m\neq-3$ and $n\neq-2,2$ if
\begin{eqnarray}\label{93}
r>r_{{\Sigma}^{(i)}}\left(\frac{7}{4-m}\right)^{\frac{1}{m+3}},
\end{eqnarray}
and
\begin{eqnarray}\label{94}
r>r_{{\Sigma}^{(i)}}\left(\frac{14+7n}{2-n}\right)^{\frac{1}{2n+3}}.
\end{eqnarray}
These two equations define the range of instability. Thus
instability of the system is subject to the consistency of
Eqs.(\ref{93}) and (\ref{94}). For $m=-3$, we obtain from
Eq.(\ref{93})
\begin{eqnarray}\label{95}
\frac{8\pi{\gamma}}{6}\left[7\log
\left(\frac{r}{r_{{\Sigma}^{(i)}}}\right)-1\right],
\end{eqnarray}
which defines the instability region for
$r>r_{{\Sigma}^{(i)}}1.15$. For $n=-2,2$, the range of instability
is not defined.

Now we find the instability range of Eqs.(\ref{93}) and (\ref{94})
for the remaining values of $m$ and $n$. For this purpose, we
consider the following two cases.\\\\
\textbf{Case (i)} Here we take $m\leq0$ and $n\leq0$. For $m=0$,
Eq.(\ref{93}) gives $r>{r_{{\Sigma}^{(i)}}}1.20$ which shows that
the region of instability decreases from $1.20$ to $1.15$ as $m$
varies from $0$ to $-3$. When $m$ varies from $-3$ to $-\infty$,
the unstable region is swept out by the whole fluid, i.e.,
$r>{r_{{\Sigma}^{(i)}}}$. For $n=0$, Eq.(\ref{94}) yields
$r>{r_{{\Sigma}^{(i)}}}1.91$. This indicates that instability
range varies from $1.91$ to $2.33$ as $n$ varies from $0$ to $-1$,
i.e.,
decreases for this region and vanishes for $n\leq-2$.\\\\
\textbf{Case (ii)} When $m\geq0$ and $n\geq0$, we see from
Eq.(\ref{93}) that the range of instability decreases as $m$
increases and vanishes for $m\geq4$, while for $n\geq0$, the range
of instability in Eq.(\ref{94}) varies from $1.91$ to $1.83$ as
$n$ varies from $0$ to $1$, i.e., the range of instability
increases as $n$ approaches to $1$ and vanishes for $n\geq2$. In
other words, electromagnetic field reduces the instability region
in the interval $(-2,2)$.\\

It is mentioned here that, for the pN approximation, the physical
behavior of the dynamical equation is essentially the same by
considering the relativistic effects upto first order
\begin{eqnarray}\nonumber
&&24\pi{\mu_0}+8\pi{\mid{p'_{r0}}\mid}r+2\left(\alpha_{\Sigma}
-\frac{21}{r^2}\right)\frac{m_0}{r}\\\nonumber&+&16\pi{\mid{p'_{r0}}\mid}m_0
+8\pi{\alpha_{\Sigma}}p_{r0}r^2+6\left(\alpha_{\Sigma}
-\frac{5}{r^2}\right)\left(\frac{m_0}{r}\right)^2\\\nonumber&+&\frac{32{\pi}^2r^3{E_0}}{k}
\left(2e+re'+re\frac{E'_0}{E_0}\right)-24\pi{\mu_0}\left(\frac{m_0}{r}
-\alpha_{\Sigma}m^2_0\right)\\\nonumber&+&64{\pi^2}E^2_0\left(1-\frac{m_0}{r}\right)
+\frac{64{\pi^2}E^2_0}{r}\left(5+\alpha_{\Sigma}r-7\frac{m_0}{r}\right)\\\nonumber
&=&32\pi(5p_{r0}-2p_{\perp0})+16{\pi}(2p_{r0}-p_{\perp0})\frac{m_0}{r}
+128{\pi^3}{\mu_0}E^2_0r(1-\alpha_{\Sigma}r^2)\\\label{98}
&+&\frac{16\pi^2}{k}r^4E_0{\mu_0}
+32{\pi^2}E_0E'_0\left(9-\frac{2m_0}{r}\right).
\end{eqnarray}

\section{Concluding Remarks}

This paper investigates the effects of electromagnetic field on the
instability range of expansionfree fluid at Newtonian and post
Newtonian regimes. In general, the instability range is defined by
the adiabatic index $\Gamma$ which measures the compressibility of
the fluid. On the other hand, in our case, the instability range
depends upon the radial profile of the energy density,
electromagnetic field and the local anisotropy of pressure at N
approximation, but independent of the adiabatic index $\Gamma$. This
means that the stiffness of the fluid at Newtonian and post
Newtonian regimes does not play any role at the instability range.
It is interesting to note that independence of $\Gamma$ requires the
expansionfree collapse (without compression of the fluid). This
shows the importance of local anisotropy, inhomogeneity energy
density and electromagnetic field in the structure formation as well
as evolution of self-gravitating objects.

We see from Eqs.(\ref{93}) and (\ref{94}) that in the absence of
electromagnetic field the region of instability is taken to the
whole fluid. However, with the inclusion of electromagnetic field,
the region of instability decreases. Thus the system is unstable in
the interval $(-2,2)$ and stable for the remaining values of $n$.
Also, Eqs.(\ref{93}) and (\ref{94}) define the instability range of
the cavity associated with the expansionfree fluid. We would like to
mention here that the unstable range will be customized differently
for different parts of the sphere as the energy density and
electromagnetic field are defined by the radial profile.

\vspace{0.5cm}

{\bf Acknowledgments}

\vspace{0.5cm}

We would like to thank the Higher Education Commission, Islamabad,
Pakistan, for its financial support through the {\it Indigenous
Ph.D. 5000 Fellowship Program Batch-VII}. One of us (MA) would like
to thank University of Education, Lahore for the study leave.

\end{document}